\documentclass[onecolumn]{autart} 

\usepackage{color,graphicx}     
\usepackage{appendix}

\usepackage{sidecap}
\usepackage{lscape}
\usepackage{setspace}
\usepackage{amsmath,amsfonts,amssymb}
\usepackage{hyperref}
\usepackage[all]{hypcap}
\usepackage[square,numbers,sort&compress]{natbib}

\begin{document}

\begin{frontmatter}
\title{Terrorism: Mechanism of Radicalization Process, Control of Contagion and Counter-Terrorist Measures}     
\thanks[footnoteinfo]{Corresponding author: A.~Cherif. Tel. +1-480-965-2115. Fax. +1-480-727-7346.}
\author[MCMSC,SHES]{Alhaji Cherif\thanksref{footnoteinfo}}\ead{alhaji.cherif@asu.edu},    
\author[PRC,EJG]{Hirotoshi Yoshioka},         
\author[IIR,SEEE]{Wei Ni},         
\author[LMSS]{Prasanta Bose}         

\address[MCMSC]{Mathematical and Computational Modeling Sciences Center, Arizona State University, Tempe AZ 85287}
\address[SHES]{School of Human Evolution and Social Change, Arizona State University, Tempe AZ 85287}
\address[PRC]{Population Research Center, University of Texas at Austin, Austin TX 78712}
\address[EJG]{Embassy of Japan in Guatemala, Guatemala}
\address[IIR]{Institute for Infocomm Research, Singapore 138632}
\address[SEEE]{School of Electrical and Electronic Engineering, Nanyang Technological University, Singapore 639798}
\address[LMSS]{Lockheed Martin Space Systems, Advanced Technology Center, Palo Alto, CA 94303}

\begin{abstract}                          
The spread of radical ideologies is key to fanaticism, recruitment and terrorist activities. Understanding the contagion dynamics of radical ideologies is fundamentally important for sociological research and counter-measures. However, theoretical understanding of social contagions such as radicalization, and integrative theories incorporating the process and its realized outcomes (e.g., terrorism, occurrence of catastrophic terrorist events) are still lacking. In this paper, we investigate the contagion process  of radicalization. We develop a stylized model that captures the radicalization mechanism (e.g., horizontal, and oblique contact process) through several stages of individuals using \textit{in silico} approach constructed from compartmentalized model. Building on the insights from epidemiology, our model entails heuristics for controlling the spread of radical ideologies on half-Gaussian social spatial networks. We find, with sufficient initial conditions, persistence or permanence phenomena. To explore the underlying mechanism of horizontal and oblique transmissions, we consider the control effects on militancy rates (e.g., recruitment into terrorist groups from radical fraternities). The effect of control exhibits a tipping point phenomena at a critical militancy rate, where the largest number of terrorists are identified and removed. The analysis suggests that the focus should be on recruitment process (e.g., which plays a major role on the dynamics of growth of terrorist groups), and on adaptive pressures that can drive organization to the critical point. Our study also shows scaling properties where the effects of different counter-terrorist identification and self-identification rates can be unfolded. That is, numbers of removed terrorists collapse on a simple pattern. This study provides a deeper understanding of adaptive malignant organizations such as \textit{Al-Qaeda}, \textit{Hezbollah}, \textit{Hamas}, \textit{Shining Path} and other extremist fraternities, while confirming other properties known in terrorism research community.
\end{abstract}

\begin{keyword}                           
radicalization mechanism, contagion, socio-spatial network, terrorism, counter-terrorism measures, sociophysics
\end{keyword}                             

\end{frontmatter}

\section{Introduction}
As a result of American-led campaign against global terrorism, specially the \textit{Salafist-Jihadist} brand, there has been an explosive scholarly research in the direction of socio-political and religious aggressive behaviors with emphasis on developing mechanism necessary for understanding and predicting attacks. The studies of terrorism have gained interdisciplinary tractions among sociologists, biologists, psychologists, anthropologists, mathematicians and physicists \cite{Chavez, Cherif, Clauset, Keohane, Pyszczynski}. Since it is difficult (if not impossible) to perform experiments as in any well-developed scientific fields, mathematical models and agent-based \textit{in silico} experiments have provided the necessary tools to study the dynamics and properties inherent in fanatic behaviors and their realized outcomes (e.g., terrorist attacks). When the tools, usually borrowed from the fields of dynamical systems and statistical physics, are coupled with data, the behaviors inherent in terrorism are highlighted. Unfortunately, current theoretical works using these tools (e.g., complex adaptive and dynamical systems) have focused on characterizing the statistics of terrorist attacks and events. Clauset and colleagues \cite{Clauset} have employed statistical physics and related methodologies to quantify the scale invariance of global terrorist attacks with emphasis on the severity of the events. Since high-quality data on terrorist group formation, recruitment process and organizational structures are scarce, it is not surprising that current theoretical works have been directed towards event dynamics and in similar line of Clauset and collaborators. In the recent decades, other models have surfaced. Castillo-Chavez and Song \cite{Chavez}, and Cherif \cite{Cherif} have modeled the processes of recruitments and radicalization using compartmental approach of theoretical epidemiology. In many ways, radicalization process resembles other social contagion processes such as rumor, opinion and idea formations and other social cascade models. These social processes have traditionally been modeled like compartmental epidemic models via \textit{mass action} assumptions, an approach also employed in this paper. However, social contagions differs greatly from epidemiological contagions. Unlike epidemics, social contagion possesses an additional transmission mechanism beyond vertical and horizontal transmission, that is oblique transmission. Oblique transmission mechanism provides another method by which ideologies can be propagated easily.  

Various scholars have acknowledge the  need for incorporating behavioral features of radicalization mechanisms in order to capture variety of recruitment processes and behaviors associated with contact processes and ideology progressions. Cherif \cite{Cherif} incorporated differential recruitment processes into an extended version of a model proposed in \cite{Chavez} and in this paper. The study of radicalization process is important in many aspects of terrorism and effective counter-terrorism. Radicalization plays a crucial role in the organizational network formations and decentralization of foot-soldiers. Besides these apparent properties of radicalization, the process is important in the counter-terrorism and de-radicalization of youth in abroad and diaspora (e.g. the United States and European countries).

Current counter-terrorism efforts have focused on either scattering, killing or capturing a terrorist organization's core leadership, reducing the threat from its central core operatives, foot-soldiers and leaders. However, terrorist organizations continue to spread at an alarming rate in many parts of the world, as a result creating various subcultures within many communities. Since September 11, 2001, the threat from radicalized  \textit{Salafist-Jihadists} has changed. Recent events in London, Madrid, Bali, and Amsterdam illustrate new trends in the involvement of local communities in terrorist activities, in which local terrorist organizations and residents utilize the memes of \textit{Al-Qaeda} as their ideological inspirations. As a result, different forms of terrorist threats have emerged in European, Canadian, and Australian cities, where most of the terrorist attempted and successful attacks such as Madrid $2004$, Amsterdam Hofstad group, London $2005$, Toronto $18$ Case, and Australia's Operation Pendennis, have been \textit{diasporic} in nature. Such a change requires a new framework to understanding, combating terrorism and/or reducing the number of possible terrorist acts. One approach is to understanding the terrorist social networks and the effect of counter-terrorist measures on the radicalization process. In the following paragraphs, we summarize previous efforts in this direction.

Recent literature has provided various mechanisms of counter-terrorism measures. Keohane and Zeckhauser \cite{Keohane} list four fundamental ways in which a state can defend its citizens against terrorist groups: 1) stock reduction; 2) flow control 3) averting actions and; 4) amelioration. Among these four ways, the first two are the most relevant in designing a strategy to control terrorist networks formation and activities. Stock reduction refers to directly reducing the stock of terror capital. For example, military action directed toward terrorist organizations or the state supporting them belongs to this category although stock reduction does not necessarily involve military action. And the second option entails ``[i]dentifying so-called `charitable organizations' with ties to terrorist groups and applying diplomatic pressure to state sponsors" to cut off the flow of funds \cite{Keohane}. Other scholars \cite{Chavez, Cherif} suggest the reduction in recruitment pool as one of the most effective mechanism in reducing the stock of terror capital.

A control strategy must entail these two components, which are related in the sense that both components require the knowledge of a targeted terrorist network. Current counter-terrorism measure sometime do not produce the intended results. Removing key leaders in a terrorist network without considering the overall organizational structure and tools produces only temporary outcomes. However, we often times lack information on such a network and even when data are available, we face a large number of missing cases for two reasons. Firstly, as compared to criminal networks such as drug trafficking networks, terrorist networks take action much less frequently, which derives from at least the following two factors: (1) the terrorist network's ability to undertake action fluctuate considerably in a short-term because it is affected by shifts in public opinion \cite{Keohane}; and (2) attacks undertaken by terrorist networks are usually ideology-oriented rather than reward-oriented. As several researchers (e.g. Pyszczynski et al. \cite{Pyszczynski}; Richardson \cite{Richardson}) posit, this is one fundamental characteristic that differentiate terrorism from criminal acts. Indeed, Richardson \cite{Richardson} states that terrorists usually act in the service of a set of their righteous ideals. Hence, terrorist acts are usually political and rarely involves psychopathology or material deprivation \cite{Turk}. When the objective is ideological, actions tend to take place less frequently because a network may wait for the right moment to act \cite{Morselli}, and/or see reduction in security alerts.

Secondly, unlike other types of criminal networks that are centralized and hierarchical, terrorist networks tend to be more decentralized \cite{Morselli,Tucker}. Therefore, identifying central players in terrorist networks that is suggested by Akbar Hussain \cite{Akbar} often does not work because of the above noted nature of such networks aggravated by possible camouflage. Besides, identifying important ties or flows is often a difficult task since ``structures with a proven level of endurance and an established reputation" including terrorist networks tend to opt for security rather than efficiency \cite{Erickson, Morselli}.

The problem of missing data is aggravated by the fact that unlike criminal networks such as drug-trafficking networks, terrorist networks, especially transnational terrorist networks such as Al Qaeda, tend to have members from quite diverse demographic backgrounds in terms of their nationality, educational status, socioeconomic status, and ethnicity \cite{Raab}. This is especially the case among today's modern terrorist networks as compared to ``traditional" terrorist organizations such as the Irish Republican Army (IRA), Ku Klux Klan (KKK) or the Kurdish Workers' Party (PKK) that tend to recruit members from a specific population \cite{Takeyh}, and are ethnically monolithic. As a result, monolithic nature of these traditional terrorist organizations induces homogeneity necessary for social cohesion of the group or subgroup.

Given the above noted characteristics, controlling terrorist networks is a cumbersome task. In this paper, following the paradigm of Castillo-Chavez and Song \cite{Chavez} and Cherif \cite{Cherif},  we model the process of the emergence and radicalization of terrorist groups, which is vital for better understanding the nature of terrorist networks. We also present a new effective \textit{in silico heuristic} mechanism of controlling terrorist networks that entails taking advantage of our knowledge about terrorist organizations albeit limited at best, and also, lessens the dilemma for authorities when dealing with terrorist organizations. That is, minimizing publicity for these organizations while preventing them from destructive acts \cite{Turk}. As will be discussed, our abstract mechanism is designed in such a way that it can be improved without transforming its fundamental structure by adding the knowledge obtain while implementing the mechanism through the update of relevant parameters using, for example, Bayesian inference. Finally, since our mechanism examines both how terrorist networks propagate and can be controlled, it will be useful for us to understand how and why individuals become terrorists (e.g., radicalization process). We develop a contagion \cite{Chavez, Cherif, Bettencourt, Bartholomew} stylized model and follow the insights from network community to investigate the dynamics of radicalization, and propose a method of effective control of such a social contagion process. As a result, our model synthesizes two approaches from network studies of terrorism and contagion processes, and allows us to bridge the two literatures on terrorism.

\section{Model Description}

This section provides the description of the proposed model and key assumptions we have made. The section is divided into four parts: network generation and characterization, radicalization mechanism, control mechanism and simulation approach. In the network section, we provide the method of generating socio-spatial network and characterization of its property. In the radicalization mechanism section, we follow mathematical and theoretical epidemiological approach of social contagion to modeling the spread or transmission mechanism and the recruitment process of radical individuals in an idealized community. To the best of our knowledge, only few papers have employed this epidemiological approach of understanding the transmission mechanism of radical and fanatic behaviors \cite{Chavez, Cherif, Santonja, Stauffer}. However, none of the previous models have investigated the impact of variability in individual traits and attributes on the mechanism of radicalization; have made distinction between radicalized and terrorist individuals; and/or proposed adaptive control mechanism necessary to defeat or reduce its organizational strength. In the last two subsections, we propose a method of control mechanism and agent-based simulation approaches, which help identify the tipping points that facilitate the radicalization process to take place and spread. In this paper, we focus primarily on individual-based (or agent-based) model adopting formalism of contagion and network to propose an effective control mechanism. In doing so, we synthesize macro-formulation of epidemiological compartmentalized model of social contagion with the heterogeneity and networked contact structures of agent-based models. The use of agent-based modeling provides opportunities for inclusion of more detailed behavioral complexities in which emergent dynamics (in large population size) are captured by the macrodynamics formulated via the differential equations.

\subsection{Network Generation and Characterization}
It has long been known that social and spatial structures are important if we are to understand the long-term persistence, and detailed emergent behaviors of social contagion process such as radicalization. However, determining a complete interaction details and/or contact network structure require knowledge of every relationship between individuals in the population of interests. As a result, it is impractical and problematic to attempt to generate contact networks requiring personal information (which are not easily available and/or volunteered). In the case of radicalization, it is not always clear (1) how to define relationships capable of permitting the spread of \text{Salafist-Jihadist} ideology; and (2) how much contacts with a radical individuals are necessary to have a measurable risk of conversion. As a consequence, the simulations of social processes (e.g., readicalization) on idealized networks (e.g., random networks; exponential-random netwoks ($p*$-models); lattices; small-world networks; spatial networks; scale-free networks) have mainly relied on observations. Therefore, we have chosen to use socio-spatial network primarily for the flexibility of generating a wide variety of network behaviors ranging from highly clustered lattices to small-world phenomena to globally connected random network arrangements with high degree of heterogeneity. Herein, we discuss the network capable of capturing transmission and recruitment processes of radicalizations, both direct (e.g., peer-to-peer) and indirect (e.g., radical sermons on the web).
\noindent
\begin{figure}[htp]\label{fig1}
\begin{center}
\includegraphics[height=9cm]{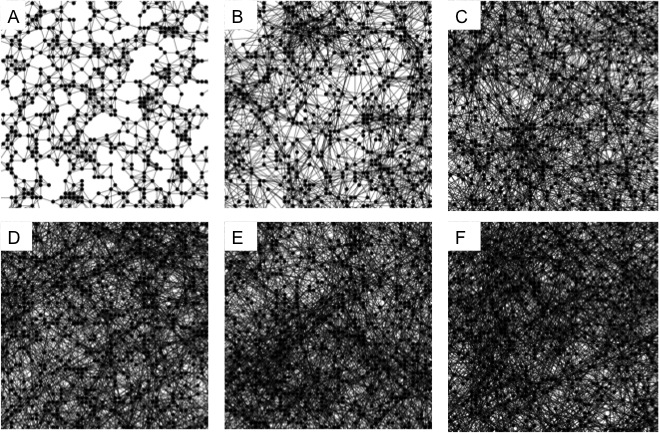}\\
\caption{The plot shows graphs for different values of $D$ [\textbf{A} ($D = 1$), \textbf{B} ($D = 5$), \textbf{C} ($D = 10$), \textbf{D} ($D = 15$), \textbf{E} ($D = 20$) and \textbf{F} ($D = 25$)]. It shows how parameter D controls not only the length of edge but also the Clustering Coefficient. On the left, the figure shows a graph for $D = 1$ (Fig. \ref{fig1}\textbf{A}) with Clustering Coefficient $= .15$, while network with $D = 10$ (Fig. \ref{fig1}\textbf{C}) has a Clustering Coefficient $= .015$.In all these graphs, we assume an average number of contacts ($<n> = 8$) and degree distributions are similar with total number of agent sent to ($N = 600$) for visual aesthetic. The average length of edges are different for all the networks. Network with $D = 1$ has smaller average length (\textit{l}) than that of network with $D = 25$ ($\textit{l}_{1} < \textit{l}_{5} < \dots < \textit{l}_{25}$). These calculations and results are due to \cite{Cohen}.}
\end{center}
\end{figure}

We use a network approach to represent the contact structure of agents in an idealized society or community. In our construction, individuals are represented by vertices with contacts between members denoted by edges. An edge represents the contact between vertices that can allow transmission of ideology. In our agent-based model, the network edges are created in such a way that a parameterized average density is achieved while maintaining the appropriate distribution. We connect two agents or nodes using a half-Gaussian spatial or socio-spatial distribution of width $D$ \cite{Keeling99, Keeling00}:
\begin{eqnarray}
p(n,D) &=& \frac{<n>}{2\pi D^{2}}e^{-\frac{d^{2}}{2D^{2}}}  \label{eqn1}
\end{eqnarray}%
where $d$ is the distance between two agents which can loosely be defined as affinity, family, friendships, neighbors, similarity, etc. The parameter $<n>$ is the average number of contacts or degree in the network, and $D$ is the spatial length-scale. 

The parameter $D$ measures the socio-spatial preferential vicinity. For instance, small $D$ exhibits local dynamics where agents will preferentially connect to other agents within their socio-spatial vicinity, while large $D$ provides global connection where agents can be connected to distant agents. In other words, the chance of a link between two individuals decreases as the distance between them increases such that ideological infection is transmitted preferentially to individuals in the proximity (whether it is locational, ideological, socio-economical or socio-psychological proximity). This construction of the network allows us to use the parameter $D$ as a proxy for clustering coefficient, which can determine transmissibility of ideology, while maintaining small contact size \cite{Keeling99, Keeling00, Keeling05, Read, Watts, Klovdahl} as shown in Fig. \ref{fig1}. For instance, in the characterizations performed by Cohen and colleagues \cite{Cohen}, they showed that lower values of $D$ yield networks with higher clustering coefficients, while higher values of $D$ produce networks with lower coefficients. Therefore, the mean length of an edge increases with $D$ while maintaining the likelihood of shorter length. Spatial networks, such as the one presented in this paper, usually exhibit a reasonably high degree of heterogeneity. For further detailed characterization of networks, the reader is advised to consult \cite{Keeling99, Keeling00, Cohen, Keeling05, Read, Watts, Klovdahl} and references therein.

\subsection{Modeling Radicalization Mechanism: Transmission on the Network}
We first derive a macro-level formulation of radicalization mechanism using tools from theoretical and mathematical epidemiology, which are later used to describe micro-interactions between agents presented in section $2.4$. We divide the population of interest, $P$ $( e.g.:P = G + C),$ into two sub-populations mainly core ($C = S + I + R + R_{s} + R_{L}$) and non-core groups ($G$). The non-core group (e.g., general population and at-risk individuals within the general population) $G$, consists of individuals who have not enter or have not come in contact with radical ideology. This stage is usually the source of recriutment pool. The core group consists of susceptible $S$, semi-radical or moderate $I$, and radical $R$ individuals who have gone through \textit{Self-identification, Indoctrinization,} and \textit{Jihadization} stages, respectively. The susceptible group includes members of the population who have not yet been converted into adapting the ideology but have begun to explore \textit{Salafi-Islamic} ideology and gradually gravitate toward radical ideas. Semi-radical includes those who have been converted and as well as those who may not be fully committed. These individuals go through indoctrination stage in which they intensify their beliefs, and reinforce their radical views. The radical group consists of individuals who have internalized extreme ideology and/or have accepted their duties as \textit{Jihadists},  and holy warriors or \textit{Mujahedeens}. It is from this subgroup that terrorist groups emerge. The radical group is further subdivided into two groups: foot-soldiers $R_{s}$ and leaders and operatives $R_{L}$.
\noindent
\begin{figure}[htp]\label{fig2}
\begin{center}
\includegraphics[height=5.5cm]{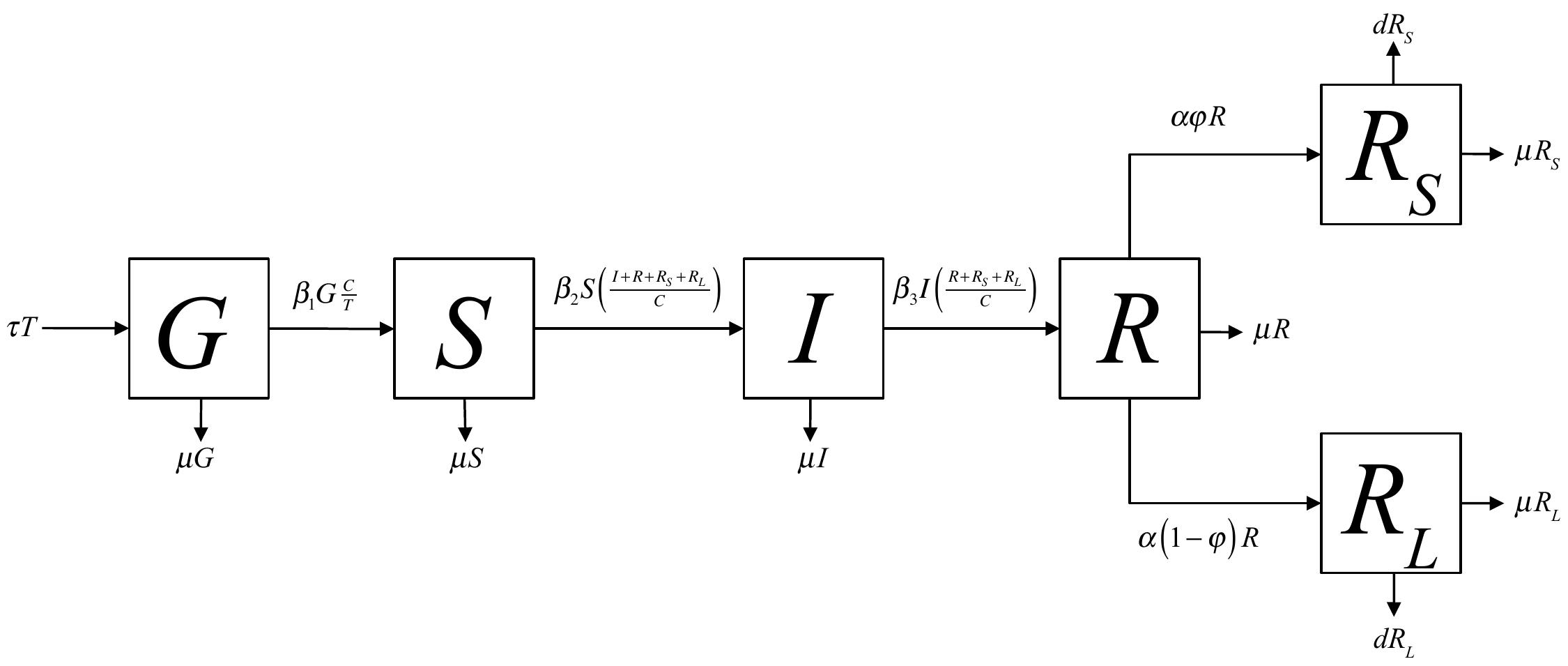} \\ 
\caption{Schematics of Radicalization Process}
\end{center}
\end{figure}
In an effort to understand the process of radicalization in the West (or in the Islamic Diasporas), Silber and Bhatt \cite{Silber} with the  help of investigators (e.g., New York Police Department detectives and analysts, intelligence officials and academics) undertook a comparative study in which they investigate ten (10) Islamic terrorist groups in the United States and Europe. In their report, they identify similar process of radicalization discussed above and in \cite{Chavez, Cherif, Comments}. They also found that four stages are consistent among all the ten groups in their comparative studies, namely pre-radicalization (e.g., noncore group, $G$), self-identification (e.g., susceptible core group, $S$), indoctrinization (e.g., semi-radical, $I$), and jihadization (e.g., radical $R$, and terrorist groups including foot-soldiers $R_{s}$ and leaders $R_{L}$). With a further analysis and study of their finding (mostly explanatory), we observe that individuals in the ten case studies (or in ten terrorist groups) spend approximately $1.66 \pm 0.844$, $3.026 \pm 1.16$, $1.91 \pm 0.899$, and $1.93 \pm 0.592$ years in at-risk general population ($G$), susceptible class ($S$), semi-radical class ($I$), and radical and terrorist groups ($R + R_{L} + R_{s}$), respectively. However, it should be noted that the time spent in these stages are increasingly becoming shorter since 2001, as the recruitment rates have considerably increased.

With the above assumptions, the ansatz describing the radicalization mechanism with horizontal transmission process shown in Fig. 2 is given as:
\begin{eqnarray}
G' & = & \tau T-\beta_{1}G\frac{C}{P}-\mu G \label{eqn2} \\
S' & = & \beta_{1}G\frac{C}{P}-\mu S-\beta_{2}S\frac{I+R+R_{s}+R_{L}}{C} \label{eqn3} \\
I' & = & \beta_{2}S\frac{I+R+R_{s}+R_{L}}{C}-\mu I-\beta_{3}I\frac{R+R_{s}+R_{L}}{C} \label{eqn4} \\
R' & = & \beta_{3}I\frac{R+R_{s}+R_{L}}{C}-\mu R-\alpha R \label{eqn5} \\
R_{s}' & = & \alpha \varphi R-\mu R_{s}-dR_{s} \label{eqn6} \\
R_{L}' & = & \alpha \left(1-\varphi \right)R-\mu R_{L}-\kappa dR_{L} \label{eqn7} \\
P & = & G + C \label{eqn8} \\
C & = & S + I + R + R_{s} + R_{L} \label{eqn9} 
\end{eqnarray}
where $\kappa \in [0,1]$. The core population recruits individuals from the non-core population at the per capita contact rate of $\beta_{1},$ which is replenished at an assumed per capita rate $\tau T$. The per capita contact rates of $\beta_{1}$ (\textit{self-identification rate}), $\beta_{2}$ (\textit{indoctrinization rate}), $\beta_{3}$ (\textit{radicalization rate}) measure the strength of the recruitment into susceptible, semi-radical and radical subpopulation, respectively. The parameter $\mu$ is the rate at which individuals leave the system, for example dying naturally, while $d$ and $\kappa d$ are the terrorist activities induced mortality or removal rates (e.g. suicide bombing, died in counter-terrorist attacks or arrests) for foot-soldiers and leaders, respectively. We assume that proportion (e.g., $\varphi \in [0, 1] $) of radicals become foot-soldiers and $1-\varphi$ proportion as leaders  at the transition rate of militancy $\alpha$. In the formulation above, we assume that promotional rate from foot-soldiers and valued operatives to leaders is neglegibly small. We further assume only horizontal and oblique transmission mechanisms, which seem to be apparent in current radicalization and self-radicalization of youth and militancy of \textit{clean-skin} individuals in Islamic diaspora. The conversion rates from self-identified to indoctrinized, and from indoctrinized to radicalized individual is given as $\beta_{2}S\frac{I + R + R_{s} + R_{L}}{C}$ and $\beta_{3}I\frac{R+R_{s}+R_{L}}{C}$, respectively.

The mathematical analyses of a model similar to the one presented in this paper have appeared elsewhere \cite{Chavez, Cherif}. However, this model differs greatly from previous models of social contagions \cite{Chavez, Bettencourt, Bartholomew, Boyd, Cavalli, Feldman, Lumsden, Santonja, Stauffer, Dennett} in that it incorporates mechanism for individuals who fall out of the recruitment process as a result of ``social funneling", and additionally includes foot-soldiers and leadership compartmental classes. In the next sections, we focus on individual-based model (or agent-based model) adopting the above formalism.

\subsection{Heuristic Agent-based Control Mechanism}

As an effort to prevent the spread of radicalization and the emergence of terrorist organizations, we propose a three-step heuristic agent-based (individual-based) control mechanism based on the above radicalization process and the recruitment structure. The three steps are namely \textit{identification}, \textit{surveillance} and \textit{action}. In the \textit{identification} step, the objective is to search and identify or encounter potential foot-soldiers and the leaders. In order to achieve economically feasible and effective control, the semi-radicals, the susceptible and the general populace are not under control. In contrast, the radicals are randomly investigated, as they are closest to turning into foot-soldiers and/or the leaders. Recent observation of resurgence of terrorist recruits in Afghanistan shows that, for example, individuals attending \textit{Madrasa} schools are easy target for \textit{Taliban} and \textit{Al-Qaeda} due to the fact that most of these individuals have already completed the first three stages of radicalization (Eqs. \ref{eqn3}-\ref{eqn5}); and share similar ideological traits. As a result, it is natural to target the radicals within this social context. In our heuristic framework of control, these three-steps are abstract construct of control mechanism. In the \textit{identification} step, radical individuals are identified as potential terrorists at a certain identification rate and efficacy, depending on the accuracy of our information about the terrorist networks or organizations. Once they have been identified, monitoring stage begins, where radicals are put under surveillance for a certain period, the length of which is flexible and depends on the efficacy of the counter-terrorist measures and resources.  Abstractly, monitoring may entail gathering information or surveillance in its literal sense. In the simulation presented in this paper, we assume that the identified radicals will be monitored for at most three years. This time window has been chosen arbitrary for the sake of simplicity and simulation speed. However, the time window can be chosen to be a stochastic process. In fact, the strength of the monitoring time constrain can be enhanced via implementation of Markovian renewal process and/or by some other mechanisms adapted from, say, optimal foraging principle in which statistical optimization objectives can be developed. For action stage, if the individuals being monitored turn into either foot-soldiers and the leaders within this surveillance period, actions are taken to remove, arrest or kill them. If this change does not occur, no actions will be taken against them and the surveillance will be tentatively over for these specific radicals. 
\noindent
\begin{figure}[htp]\label{fig3}
\begin{center}
\includegraphics[width=5.5in]{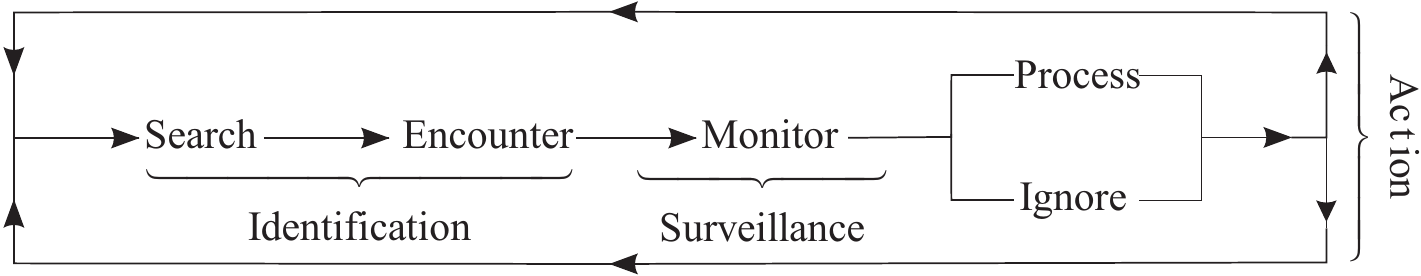}\\ 
\caption{Schematics summarizing the control mechanism}
\end{center}
\end{figure}

\subsection{Individual-Based Model (IBM) Simulations}

In the proposed model, we use macro-level dynamics (Eqs. $2$-$7$) to derive and implement appropriate micro-interaction dynamics between agents using networked social spatial contacts generate on half-Gaussian (Eq. \ref{eqn1}). Each individual is given a set of distributed attributes (e.g. age, education, desperation, income level), and we define individual propensity, $\theta_{i} = \sum_{j = 1}^{|A|} \rho_{j} \theta_{ij}$, as a convex combination (where $|A|$ is the number of attributes and $\sum_{j = 1}^{|A|} \rho_{j} = 1$) of its attributes. We use convex combination assumption because of difficulty associated with understanding which social experiences, interactions and conditions that increase the likelihood of becoming radicals, and because of conflicting and sometime contradictory explanations and accounts. The existence of cohesively homogenous factors and ethnically \textit{monolithic} with hierarchical structures that resemble corporation or military organizations are too simplistic to explain the recent varieties in various modern transnational terrorist fraternities. For example, socio-economic factors or educational status alone are not sufficient and necessary condition for suicide bombers involved in September $11$ and other Western terrorist threats.

\begin{figure}[htp]\label{fig4}
\begin{center}
\includegraphics[width=7.1in]{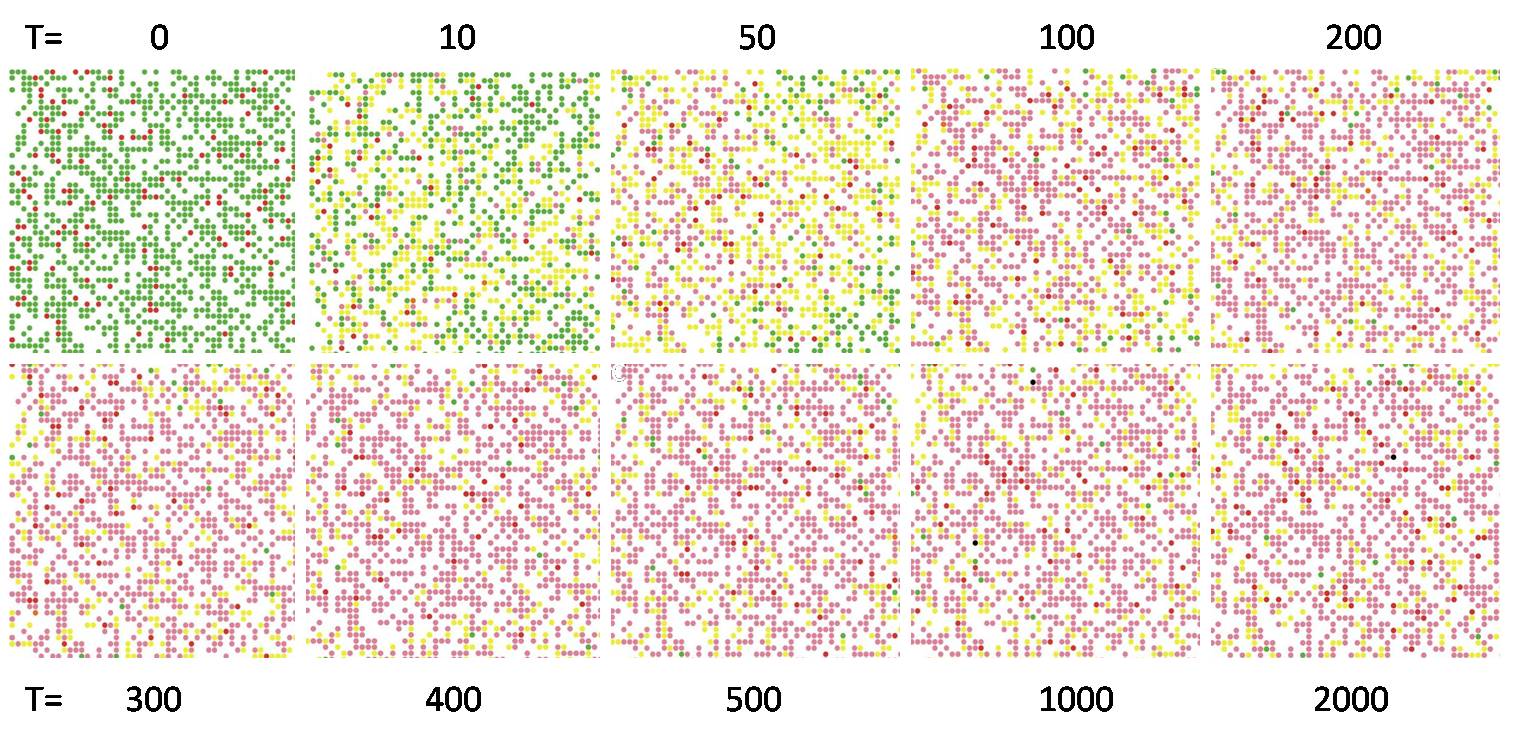}\\ 
\caption{Shows evolution of radicalization process for parameter values for terror-induced mortality ($d=0.5$, $\kappa = 1$), natural death rate or mortality ($0.2$), number of agents ($1250$ for visualization purpose), initial number of radicals ($10\%$ of total population), \textit{indoctrinization rate} ($0.25$), average degree node ($<n> = 10$), \textit{Self-Identification rate} ($0.2$), \textit{Jihadization rate}($0.5$), \textit{Militancy rate} or progression ($0.2$), control efficacy ($0.5$) and $D=10$. Links have been hidden for visualization purpose, and here T denotes time.}
\end{center}
\end{figure}

For transmission of radical ideology, we define ideological transmissibility $\tau_{i}$ of each agent as a normalized propensity $\theta_{i}$. The transmissibility $\tau_{i}$ of $i^{th}$ agent is given as:
\begin{equation}
\tau_{i} = \frac{\theta_{i}}{\sum_{i=1}^{N}\theta_{i}} \label{eqn8}
\end{equation}
where $N$ is total number of agents. We assume that many of the formed opinions, and observed and realized behavioral dynamics are rooted in the social structures or contexts to which they belong. As a result, an individual enters the susceptible class with individual transmissibility $\tau_{i}$ for each month that s$/$he is in contact with radicalized individuals. For example, a susceptible begins radicalization process with probability $p_{k} = 1-(1-\tau_{i})^{k}$, where $k$ is the number of radicalized contacts in a given month. We assume that individuals who have been influenced by radical ideology for the first time transition to \textit{indoctrinized} class. Similar probability is generated for progressions from \textit{indoctrinized} class to radical group. At any given time, individuals can leave the class at a rate of $\mu$. For simplicity, individuals who leave the system, at $\mu$, $d$ and $\kappa d$, are automatically reintroduced into the system but with different links for consistency and in order to maintain constant total population, making the half-Gaussian socio-spatial network a quasi-static. In our formulation, we assume the network to be static (e.g., quasi-static in the sense that individuals who leave the system or die are automatically re-introduced with different links) primarily due to the three properties associated with radicalization process. That is, the spread or contagion of ideology is extremely fast (e.g., faster time scale) as compared to changes in the network and contact structures. In addition, horizontal (e.g., peer-to-perr, face-to-face, or interpersonal social influence) and obligue (e.g., political and/or ideological pronouncement and exhortations, radical sermons posted on web, newspapers, radio and television, and books) transmissions make slow dynamic networks behave quasi-statically. Therefore, it is reasonable to assume a quasi-static contact networks with $D$ serving as preferential vicinity parameter. We assume that only active radicals can be militants at a militancy rate $\alpha$, where fractions $\varphi$ and $1-\varphi$ become foot-soldiers and leaders, respectively.

\section{Results and Discussion}

Previous military-based counter-terrorist measures have focused primarily on eliminating foot-soldiers and/or core leaders. However, because of decentralized nature of most terrorist organization, these measures have proven unsuccessful. Current strategies have recently focused, in addition to military strategies, on diminishing the strength and ability of Al-Qaeda and its affliates to recruit new members, which serve as resource for carrying terrorist attacks and operations. From the mean-field macro-behaviors of the aggregated dynamics (Eqs. \ref{eqn2}-\ref{eqn7}), the mechanism of radicalization illuminates the importance of recruitment on the spread of fanatic ideologies, and shows four functional thresholds. That is, $(i)$ $R_{j}=\frac{\beta_{j}}{\mu},$ $j=1,2$; $(ii)$ $R_{3}=\frac{\beta_{3}}{\mu + \alpha}$; and $(iii)$ $R_{T}(\alpha,d)=R_{3} [1+\frac{\alpha \varphi}{\mu + d}+\frac{\alpha (1-\varphi)}{\mu + \kappa d}]$. 

In order to sustain the core group ($C$) and to have recruitment pool for terror stocks, and memberships, the threshold $R_{1} \geq 1$ must be satisfied. The threshold $R_{1} \geq 1$ is a necessary condition for establishment of core group. Since there are always recruitment pool in the general population or non-core ($G$), the core population can be established whenever the rates of \textit{self-identification} or \textit{residence time} $\frac{1}{\mu}$ are large or long enough, respectively. Therefore, $R_{1}$ measures the strength of recruitments into core population from noncore or general population. However, to maintain radical groups and terrorist, $R_{1} \geq 1$ and $R_{2} \geq 1$, and $R_{3} \geq 1$ must be satisfied, respectively. Unlike the model proposed in \cite{Chavez}, $R_{3} \geq 1$ is not a sufficient condition for the establishment of radicalized terrorist groups, highlighting the funneling property of radicalization and recruitment processes. It is, however, necessary for the establishment of non-terrorist radical groups. For example, $R_{3} < 1$ is required for the collapsed of radical groups (non-terrorist group). However, Figs. $4$-$5$ suggest that sufficient initial number of radical members can drive the core population to radicalism before extinction, suggesting the existence of some intrinsic subtleties in the process of radicalization and the complexity of social contact of contagion processes. 

These intrinsic subtleties necessitate the need for an additional threshold for proper quantification of terrorism and radicalization processes. The threshold for terrorism, $R_{T} \geq R_{3} \geq1$ and $\alpha > 0$ is required for persistence of terrorist group. Note that $R_{T} (0,d) = R_{T} (\alpha,0) = R_{T} (0,0)=\frac{\beta_{3}}{\mu} \geq R_{3}$. Whenever, $R_{3} \geq 1$, as we increase the rate of militancy or progression from radical to terrorist ($\alpha > 0$), the number of terrorist members increases (see \ref{fig6}), suggesting that different recruitment structure requires different counter-terrorist measure designed to mitigate, hinder and/ or halt organizational strength of terrorist organizations. It should be noted that the conditions $(i)-(ii)$ are similar to the results found in \cite{Chavez, Cherif}. That is, in the absence of $\alpha$ in $R_{3}$ ($e.g., \alpha = 0$), the conditions and dynamics are equivalent to the thresholds and behaviors obtained in \cite{Chavez}. However, the mechanism that generates the hysteresis effects observed in epidemiological models of social contagion processes as results of linear (e.g., see \cite{Chavez}) and nonlinear (e.g., see \cite{Mubayi, Song, Cintron}) signatures of relapse mechanisms is absent from our model's formulation. 

From the \textit{in silico} parameter space experiments (Fig. \ref{fig6}-\ref{fig7}), the difficulty of defeating terrorist organizations such as Al-Qaeda under previously adopted counter-terrorist measures is highlighted. Whenever the recruitment methodologies are not hindered, increase in the rate of militancy increases the number of terrorist members, specially in the number of foot-soldiers as shown in Fig. $5$. Plots in Fig. \ref{fig6}(Left)  show the effectiveness of counter-terrorist measures. As identification rate of control varies, we observe its impact on the magnitude of the control efficacy (a quantity of the percent of removed radicals under counter-terrorist measures) with two distinct behaviors. For instance, an increase in self-identification rate with a fixed militancy rate induces a linear relationships with regard to control efficacy. If the self-identification is instead assumed to be fixed and the rate of militancy varies, the efficacy exhibits saturation and \textit{tipping point} phenomena. That is, the efficacy increases initially to a critical value before slightly decreasing and remaining constant. The tipping point occurs on the parameter space of militancy rate ($\alpha$) at a critical value of $\alpha_{c} = 0.2$ for all other parameter values (e.g. counter-terrorism \textit{identification rate}, \textit{self-identification rate}, and other parameter values) assuming the values in Fig. \ref{fig6}. 
\noindent
\begin{center}\label{fig5}
\begin{figure}[htp]
\begin{minipage}[b]{0.5\linewidth} 
\centering
\includegraphics[width=8.5cm]{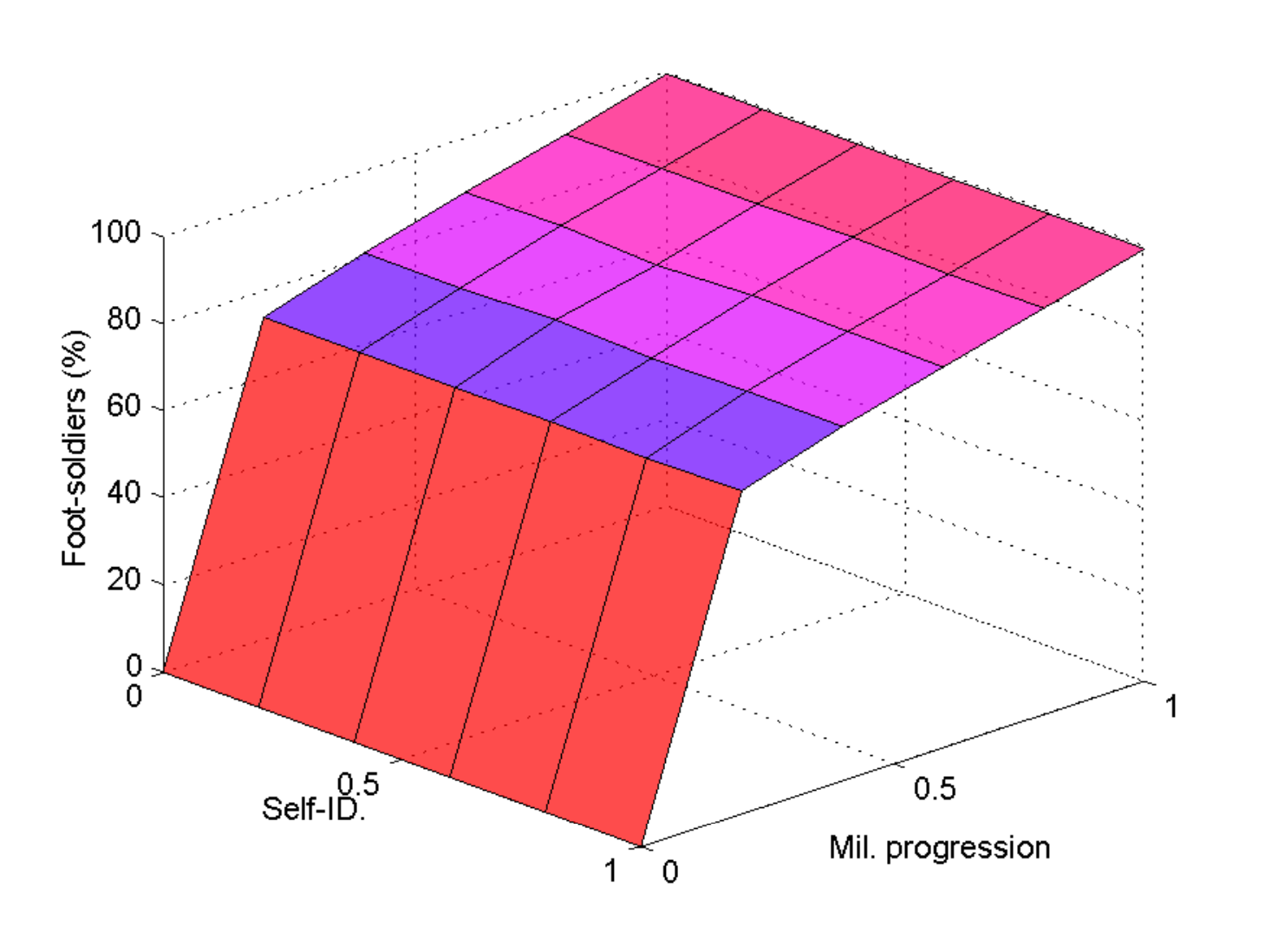}
\end{minipage}
\hspace{0.2cm} 
\begin{minipage}[b]{0.5\linewidth} 
\centering
\includegraphics[width=8.5cm]{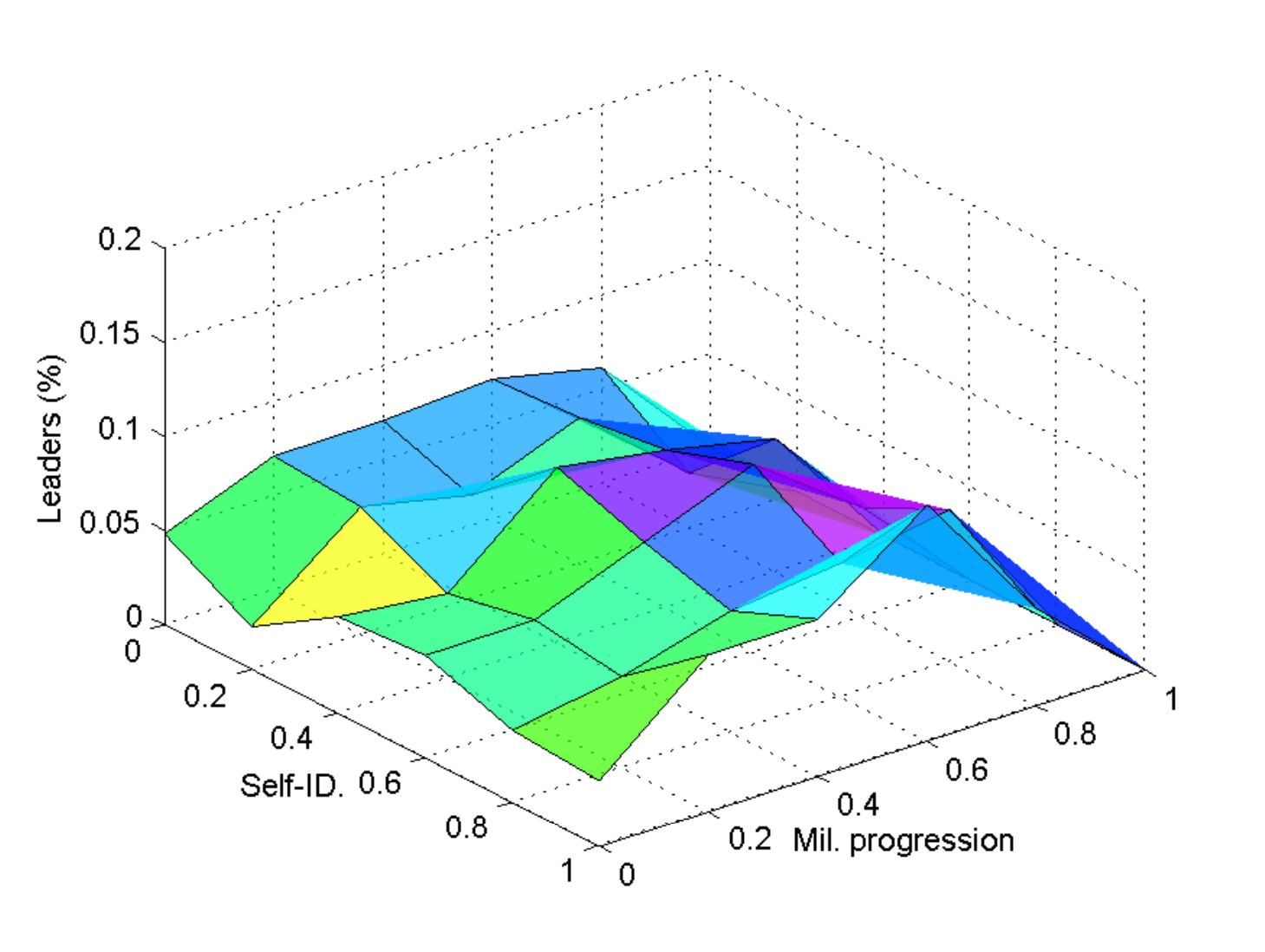}
\end{minipage}
\caption{\textit{In Silico} parameter space Analysis. Plots shows the proportion of radicals (e.g., radicals, foot-soldiers, and leaders) once a steady state has been reach (e.g., $T=2400$). Same parameter values as in figure 3, except the control efficacy, control identification rate are set to zero, and rates of \textit{Self-identification} and \textit{Militancy} are variable.}
\end{figure}
\end{center}
Figure \ref{fig6}(Left) shows the effect of counter-terrorist identification rate on the efficacy of the control or on the magnitude. However, a closer look at the dynamics suggests that the impact of counter-terrorism measures, approximated by identification rate, can be removed or unfolded. Plots on the right of Fig. \ref{fig6} show that, after a removal of the effects of counter-terrorist (CT) measures, values of percent of removed terrorist (normalized) for each \textit{self-identification} rate are relatively the same for different values of \textit{CT identification} rates. That is, the values for percent of removed terrorist as a function of the rate of militancy for \textit{self-identification} rate of $0.2$, for example, are comparably the same for all counter-terrorist identifications (i.e., $0.1$, $0.6$, $1.0$) (see Fig. \ref{fig5}). These figures (\ref{fig6}-\ref{fig7}) also reveal some of the intrinsic characteristics of radicalization and recruitment processes. One of the highlighted properties revealed from the different scaling factors used to renormalized the percent of removed terrorists is that as the rate of militancy increases, the ability of all size-fit counter-terrorism measures will not effectively mitigate the impacts of terrorism; and will not weaken the organizational ability of terrorists to act. 
 \noindent
\begin{center}
\begin{figure}[htp]\label{fig6}
\begin{minipage}[b]{.5\linewidth} 
\centering
\includegraphics[width=8.6cm]{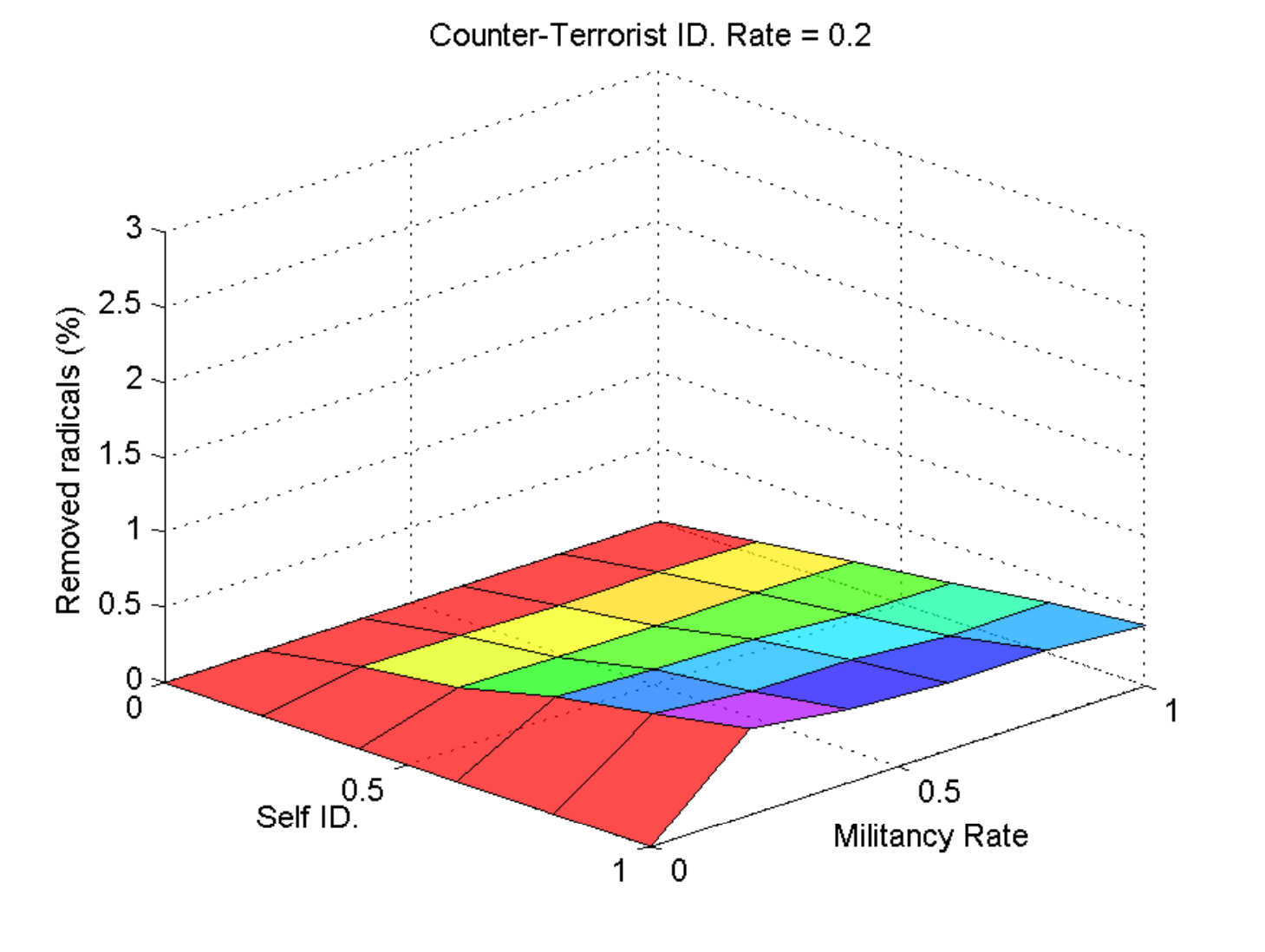}
\end{minipage}
\hspace{0.2cm} 
\begin{minipage}[b]{.5\linewidth} 
\centering
\includegraphics[width=8.6cm]{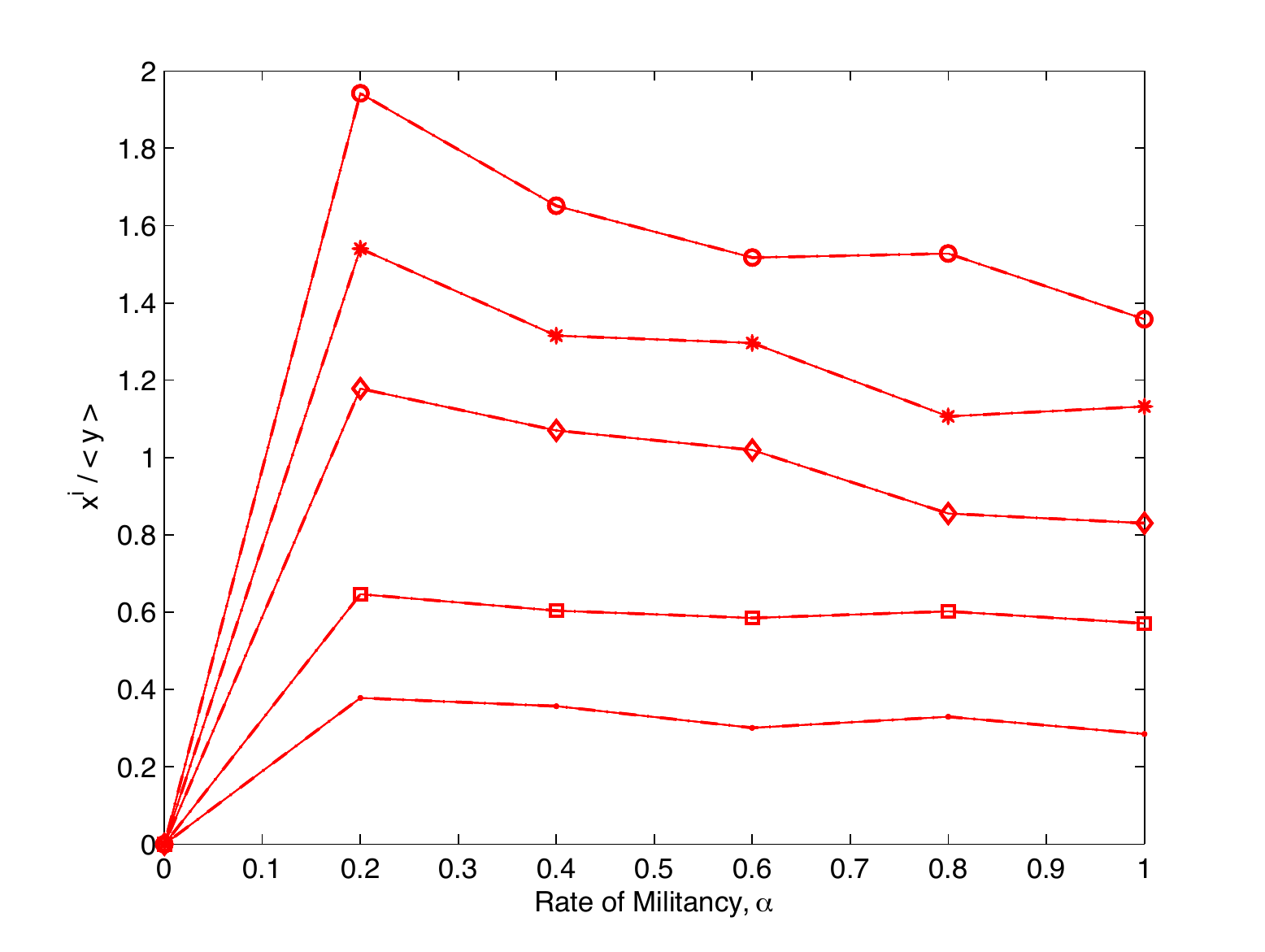}
\end{minipage}
\hspace{0.2cm} 
\begin{minipage}[b]{.5\linewidth} 
\centering
\includegraphics[width=8.6cm]{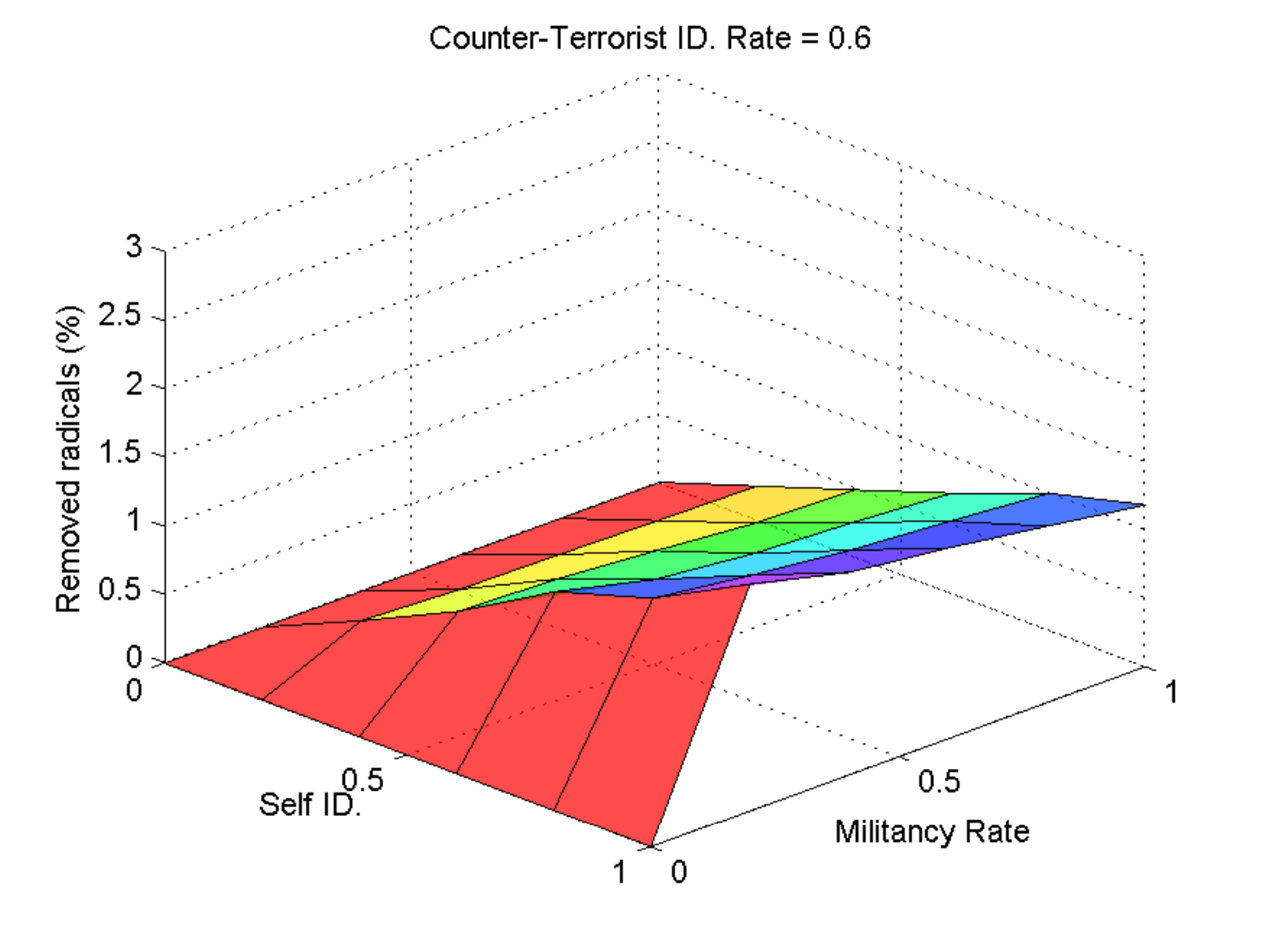}
\end{minipage}
\hspace{0.2cm} 
\begin{minipage}[b]{.5\linewidth} 
\centering
\includegraphics[width=8.6cm]{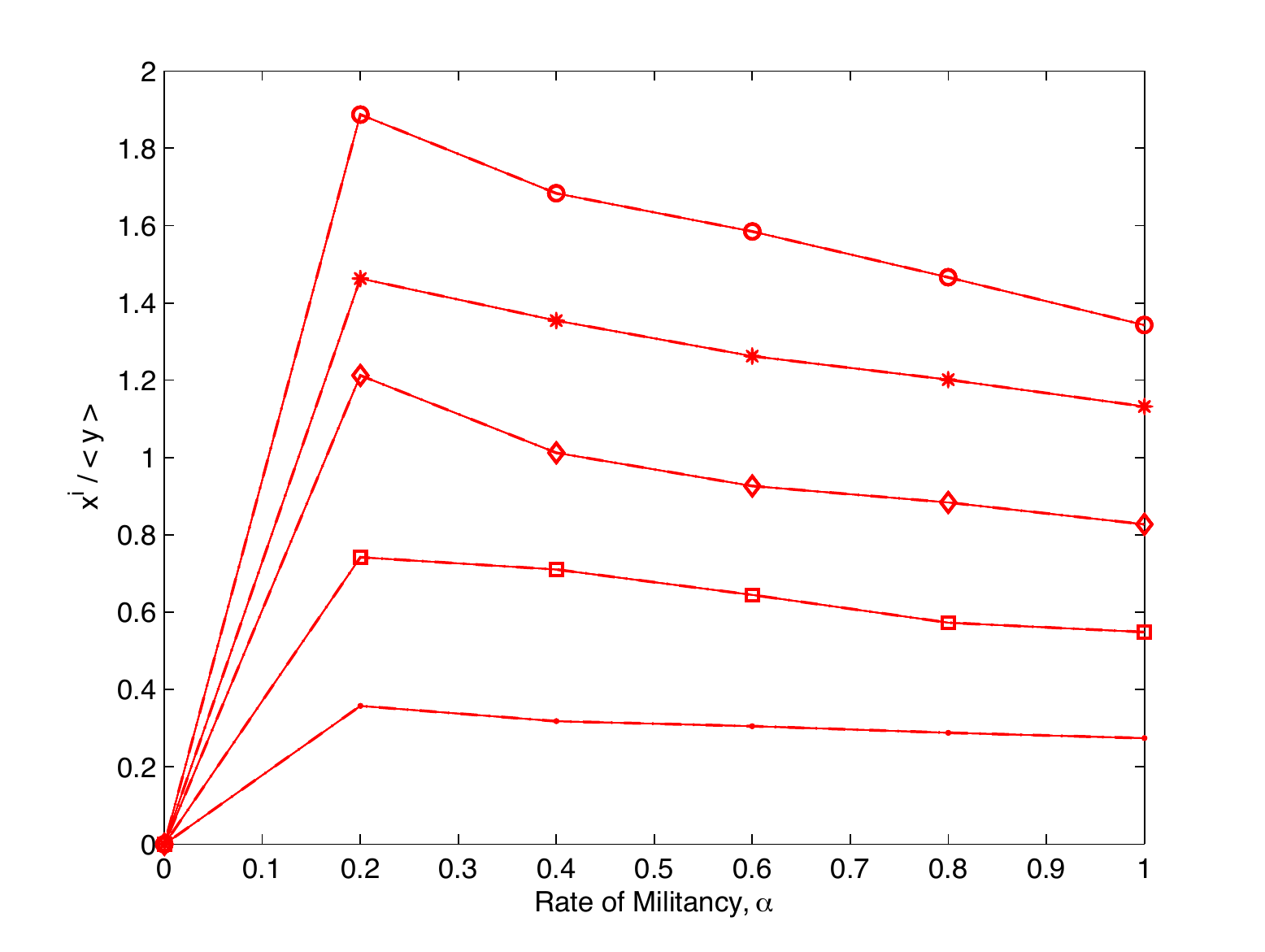}
\end{minipage}
\hspace{0.2cm} 
\begin{minipage}[b]{.5\linewidth} 
\centering
\includegraphics[width=8.6cm]{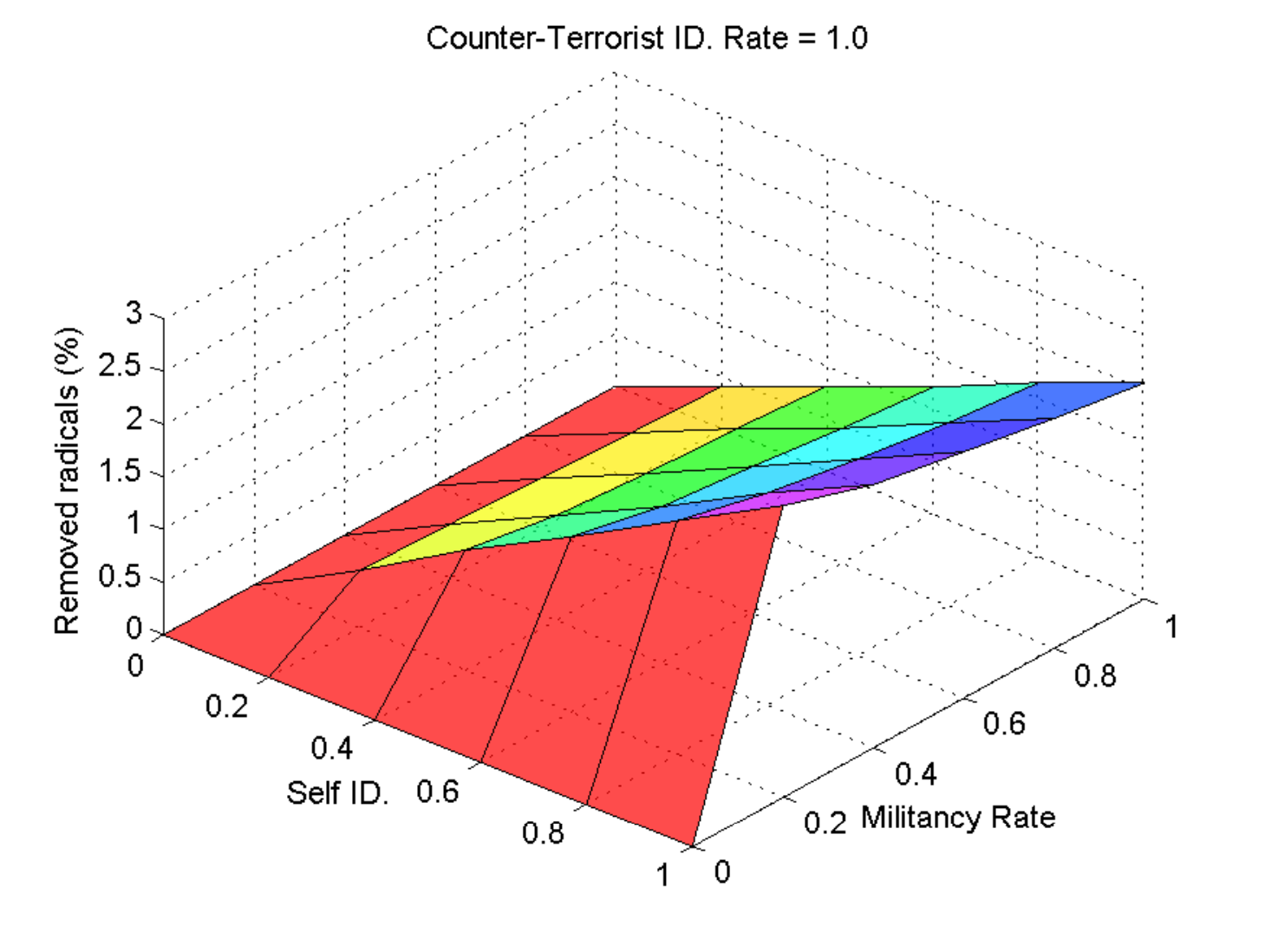}
\end{minipage}
\hspace{0.2cm} 
\begin{minipage}[b]{.5\linewidth} 
\centering
\includegraphics[width=8.6cm]{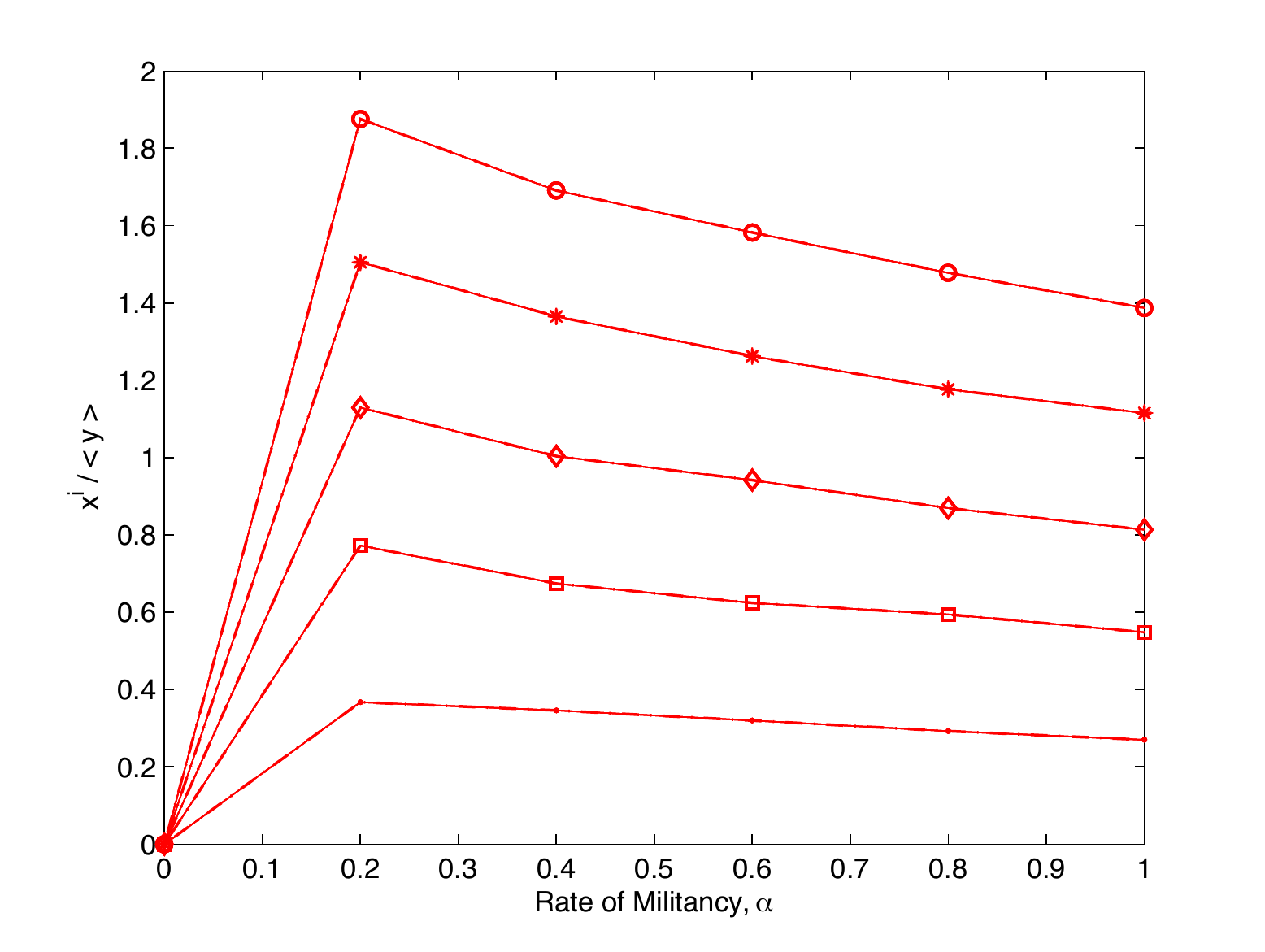}
\end{minipage}
\caption{The plots on the left show the impact of counter-terrorism \textit{identification} rate of control mechanism on the magnitude of percent on removed terrorists (e.g., proxy control efficacy) as a function of \textit{Militancy} (or Mil. progression) and \textit{Self-identification} rates. On the right, the plots show the unfolding of the plots on the left, where unfolding removes the effects of counter-terrorism identification rates and leaves the effect of self-identification rates on the percent of removed terrorists. $<y> = \frac{1}{N_{d}} \frac{1}{N_{d}} \sum_{i}^{N_{d}} \sum_{j}^{N_{d}} <x_{(ij)}> $, and $x^{(i)}$ is the percent of removed terrorist, where $i =1,2,...,5$ for \textit{Self-identification} rates of $0.2$ ($\bullet$), $0.4$ ($\square$), $0.6$ ($\Diamond$), $0.8$ ($\ast$), and $1.0$ ($\circ$), respectively.}
\end{figure}
\end{center}
Despite the fact that there is a tipping phenomena at militancy rate $\alpha_{c} = 0.2$ where the percent of removed terrorists is at its highest (Fig. \ref{fig6}), counter-terrorism measures need to be adaptive as the rate of militancy increases, suggesting that different militancy rates require different or hybrid of different control measures (e.g., Surge in Iraq, which required policy change, and new strategies including counter-insurgencies). The changes in the numbers of removed radicalized terrorists as militancy rates increase is the result of two countervailing and counter-interacting factors. The first factor is the militancy rate itself. As it increases, the more likely that radical individuals become terrorists and the greater the potential for terrorist activities and threats. But the quantity of radicalized terrorists depends on the radicalization mechanism (which exhibits hysteresis effect). The second factor is the counter-terrorist measures. As the counter-terrorist identification rates increase, the more likely it is that a terrorist will be identified. As results of these factors, our \textit{in silico} parameter space analysis suggests that it is impossible to eradicate terrorism completely when the rate of militancy and/ or recruitment efforts of fanaticism are substantially high. In fact, the results provide insight into why some terrorist organizations collapsed or are weakened (e.g., \textit{Sendero Luminoso} or Shining Path), while others are not (e.g., \textit{Hezbollah} in Lebanon) under similar counter-terrorist measures. In the remaining paragraphs of this section, we outline some implications and insight into the dynamics and behaviors of some recent terrorist organizations (e.g., \textit{Shining Path}, \textit{Hezbollah} and \textit{Al-Qaeda})

Our simulation suggests that, in the case of Shining Path, the recruitment sources were substantially diminished prior to its collapse. The recruitment pool of Shining Path was reduced, for instance, in two ways: the organization imposed taxes on their supporters (farmers), and expanded into urban areas where their supports were minimal. Because of these and its organizational structure where no clear promotional mechanism was put in place, the organization collapsed when its leader, Abimael Guzman, was arrested in 1992. As a result, the strength of the organization has diminished even though some members are actively engaged in sporadic terrorist and drug trafficking activities. Unlike Shining Path, \textit{Hezbollah} on the other has persisted despite of similar counter-terrorist measures taken by Israel over the years. In contrast to Shining Path, \textit{Hezbollah} has clear promotional mechanism; has not expanded their objectives and reach beyond their regions of interests; and has continuously tried to maintain its recruitment pools through various measures (e.g. charity, social welfare, education).
\noindent 
\begin{figure}[htp]\label{fig7}
\begin{center}
\includegraphics[width=12cm]{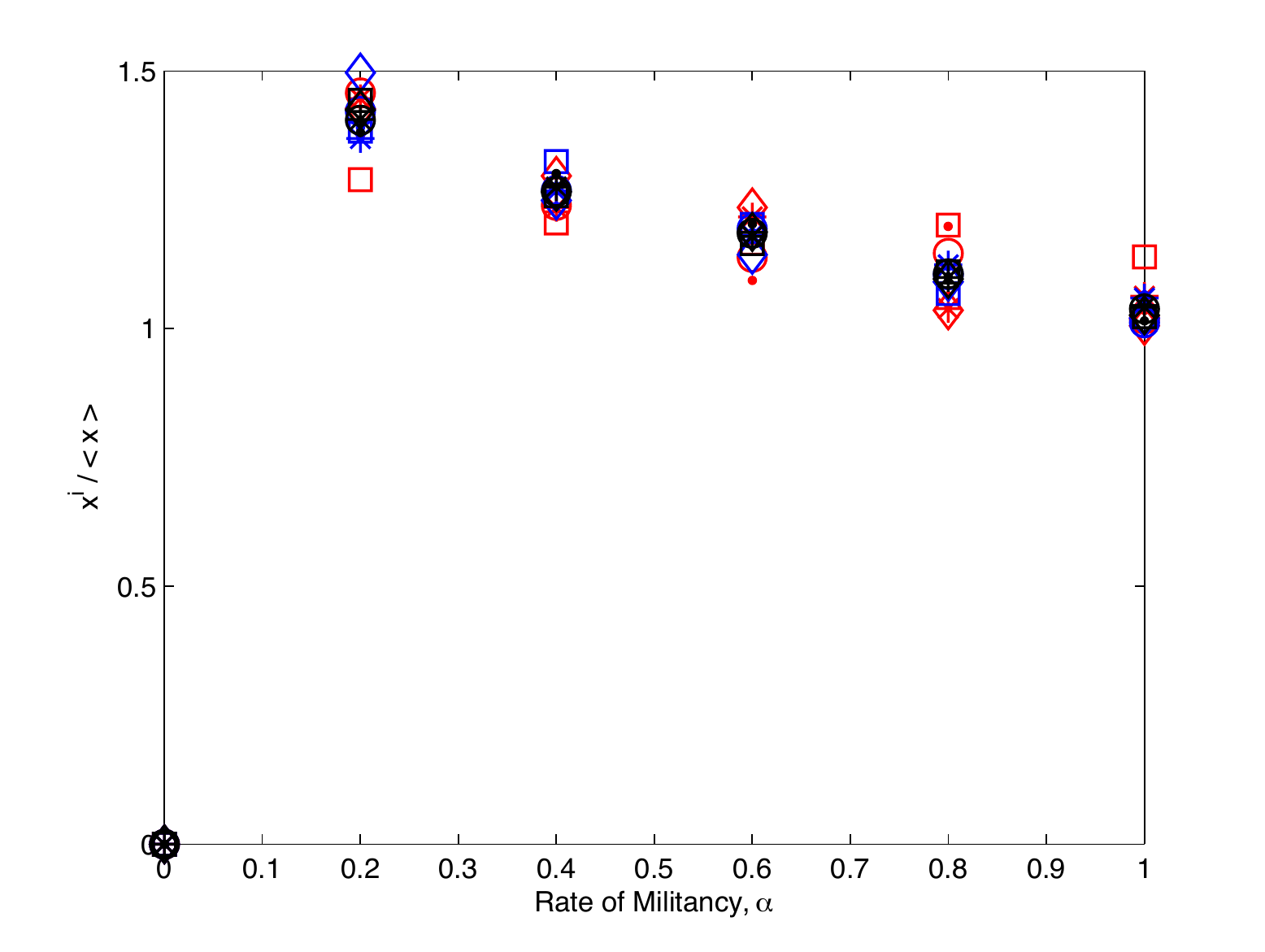}
\caption{Shows how the effect of counter-terrorism \textit{identification} and \textit{self-identification} rates on the normalized percent of removed terrorists due to counter-terrorism measures. $<x_{i}> = <x> = \frac{1}{N_{d}} \sum_{j}^{N_{d}} <x_{ij}> $, and $x^{(i)}$ is the percent of removed terrorists. The legends as follows: for self-identification rates of $0.2$ ($\bullet$), $0.4$ ($\square$), $0.6$ ($\Diamond$), $0.8$ ($\ast$), and $1.0$ ($\circ$); and the colors represent the control identification rates $0.2$ (red), $0.6$ (blue) and $1.0$ (black), respectively.}
\end{center}
\end{figure}

In term of counter-terrorist measures, the counter-terrorism policy of Israel was not effective and efficient in defeating \textit{Hezbollah}. The measures focus on targeting the leadership of \textit{Hezbollah}. Therefore, from our simulations, terrorist organizations can only be defeating \textit{if and only if} there is a decline in the recruitment pools and organizational strengths, and the organizations have weaker promotional mechanisms in place. In other words, it is impossible to defeat such terrorist organizations without reducing the strength of the organization through reduction in the number of foot-soldiers and leaders, and forcing the organization to adopt weaker leadership promotion mechanism. If a terrorist organization has a weaker promotional mechanism, then counter-terrorism measures that target the core leaderships may produce the desire outcomes (e.g. disintegration, collapse of the organization). This was the case in the aforementioned collapse of Shining Path, due in part to its hierarchical structure like most criminal organizations. If targeting leaders fail in hierarchical organization, it usual produces or creates factions within the organization, hence reducing its strength and effectiveness because of both inter- and intra-organizational conflicts. However, this approach does not work well on organizations with stronger promotional mechanism and highly decentralized recruitment pool like \textit{Hezbollah} and \textit{Al-Qaeda}. Unlike criminal fraternities with hierarchical organizational structures, modern terrorist organization like \textit{Hezbollah, Hamas} and \textit{Al-Qaeda} are decentralized and less hierarchical. As a result, these organizations are resilient to counter-terrorism actions focusing primarily on removal of key leaders. 

One interesting feature of \textit{Al-Qaeda} that makes the organization more resilient to some of the previously proposed counter-terrorist measures is its hybrid structure. As an organization, \textit{Al-Qaeda} has used various successful properties and strategies from previous traditional terrorist organizations such as the Irish republican army (IRA), Kurdish Worker Party (PKK) and the Basque separatist group ETA who developed complicated cell networks to prevent infiltration and monitoring by various security agencies. \textit{Al-Qaeda} as an organization has hierarchical structure at its leadership level, and highly decentralized networks of foot-soldiers with minimal hubs and set of satellite cells which may comprise one or more members \cite{Krebs}. Unlike the IRA, PKK and ETA, \textit{Al-Qaeda} has no geographical, national, educational, socio-economic, ethnic and socio-psychological limiting characteristic drivers usually observed in the traditional terrorist organizations and highly sophisticated criminal fraternities. This hybridization of various structures with horizontal, vertical and oblique transmission of its \textit{Salafist-Jihadist} memes allows the organization to be able to carry out stylized hierarchically coordinated attack in September 11, 2001, Iraq and Afghanistan, while minimizing their vulnerability to surveillance and infiltrations. For example, infiltrating one cell may not prevent the would-be attack by an alternative cell or reveal its intention or organizational structure. As noted by Krebs \cite{Krebs}, such a hybrid structure trades efficiency for secrecy and avoidance of detention. Because of the non-hub like structure of \textit{Al-Qaeda} foot-soldiers and operatives, we emphasize the focus of counter-terrorist measures should target emergent peripheral sympathizers and new recruits due to the fact that previous surveillance activities have usually tracked the long-term terrorists instead of new members and ``clean-skinned" lower rank members. As previously mentioned in this paper, the most effective counter-terrorist measures should be capable of forcing the organization of interest to adopt structures that might be vulnerable to further measures. For example, forcing \textit{Al-Qaeda} to purely adopt hierarchical structure due to reduction in its human resources (new members and recruits) and other pressures, a progressive removal of its leaderships might eventually lead to its defeat and demise as a terrorist organization. 

\section{Conclusion}
In this paper, we have investigated the dynamics of radicalization process on socio-spatial networks with short and long range interactions parametrized by $D$. The proposed individual-based model exploits social contagion mechanism based on epidemiological model, in which individuals were assigned a distributed convex set of attributes. The proposed model provides simplified descriptions of the intrinsics of radicalization that produce realized and emergent social dynamics of terrorism, but can be augmented with an inclusion of individual learning rubrics. Our formalism allows us to include socio-economic, psycho-pathological and individual attributes in order to produce heterogeneity in the population. Studying radicalization process from complex system theory and socio-physical perspectives provide a meaningful approach to understand the characteristics (e.g., robustness, adaptivity, autonomy, distributed functioning) of emergent networks and terrorist social organizations; and to discover the optimal pattern of beliefs, attitudes and behaviors in the hope that the adaptivity of terrorist organizations can also highlight its edge of vulnerability and sub-optimality. These, in turn, will reveal the controllability and \textit{capturability} conditions that are necessary for designing effective adaptive and \textit{reactive} counter-terrorist control measures. The macrodynamics described by coarse-grained observations of the \textit{in silico} simulations which have captured both the classical results (eqs. \ref{eqn2}-\ref{eqn9}) and some exotic behaviors that are not observable in the deterministic results. This is primarily due to the fact that the parameters $\beta_{i}$'s are country and context depended, and may depend nonlinearly on individual attributes (inclusion of which are visible in agent-based models). As a result, the use of both approaches are complementary in our understanding of radicalization process and recruitment mechanism studied herein. 

Our analysis of the parameter space revealed simplicities in the effect of militancy amidst all the complexities of radicalization mechanism and recruitment processes. We not only find tipping point and permanence phenomena, but we observed that numbers of removed radicalized terrorists due to counter-terrorist measures can be unfolded and collapsed onto a simple functional shape for all self-identification and CT identification rates. The tipping point was found to be at a critical value of $\alpha_{c} = 0.20$. Theoretical understanding and integrative theories of radicalization mechanism and recruitment process can benefit from coupling and synthesis of models like the one proposed in this paper with data. As previously alluded to, studies beyond statistical characterizations of events can provide methodologies to investigate social contagions and to propose effective measures of controls (e.g., de-radicalization, counter-terrorist measures, counter-insurgency policies). For future work, it will be interesting to investigate the impacts and dynamics of both vertical (transmission within family-members); to study rapidly evolving traits of terrorist organizations and design strategies that put adaptive pressures on them. Another direction of further study will be to develop approaches and/or models capable of combine current events, and radicalization mechanism and recruitment models to estimate relevant parameters. Synthesis and analysis of data-driven agent-based models and complex adaptive dynamical systems will become a powerful tool for the scientific study of social contagions and behaviors in the society. 

\begin{ack}
This work was completed while the authors were participating in the Complex Systems Summer School, organized by Santa Fe Institute (SFI) in June 2009. We also thank Dr. Tom Carter for his help on Netlogo. One of the authors (A.C.) is supported in part by National Science Foundation (NSF) under LSAMP Bridge to Doctoral Program Fellowship, Sloan Graduate Fellowship and Arizona State University Graduate College Diversity Enrichment Fellowship; and NSF grant DMS 0502349, Sloan Foundation and the National Security Agency (DODH982300710096) under the auspice of Mathematical Computational Modeling Sciences Center (MCMSC). The authors also thank the three anonymous reviewers, Dr. Marco Janssen and Kamal Barley for their comments and suggestions to improve the manuscript.
\end{ack}

\end{document}